\documentclass[12pt]{article}
\usepackage[dvips]{graphicx}
\usepackage{amssymb}
\usepackage{amsmath}
\usepackage{eufrak}
\newcommand{\bi}{\bibitem}
\newcommand{\bea}{\begin{equation}}
\newcommand{\eea}{\end{equation}}
\newcommand{\be}{\begin{eqnarray}}
\newcommand{\ee}{\end{eqnarray}}
\newcommand{\nn}{\nonumber}

\def\hbar#1{\backslash\hspace{-2mm}#1}

\def\nn{\nonumber}

\catcode`\@=11
\def\lsim{\mathrel{\mathpalette\@versim<}}
\def\gsim{\mathrel{\mathpalette\@versim>}}
\def\@versim#1#2{\vcenter{\offinterlineskip
\ialign{$\m@th#1\hfil##\hfil$\crcr#2\crcr\sim\crcr } }}
\catcode`\@=12

\def\2tvec#1#2{
\left(
\begin{array}{c}
#1  \\
#2  \\
\end{array}
\right)}

\def\mat2#1#2#3#4{
\left(
\begin{array}{cc}
#1 & #2 \\
#3 & #4 \\
\end{array}
\right) }

\def\Mat3#1#2#3#4#5#6#7#8#9{
\left(
\begin{array}{ccc}
#1 & #2 & #3 \\
#4 & #5 & #6 \\
#7 & #8 & #9 \\
\end{array}
\right) }

\def\3tvec#1#2#3{
\left(
\begin{array}{c}
#1  \\
#2  \\
#3  \\
\end{array}
\right)}

\def\hbar#1{\backslash\hspace{-2mm}#1}

\numberwithin{equation}{section}

\begin{document}

\begin{titlepage}
\begin{flushright}
KANAZAWA-11-08
\end{flushright}
\begin{center}

\vspace{1cm}
{\large\bf Direct and Indirect Detection of Dark Matter \\
in $D_6$ Flavor Symmetric Model}
\vspace{1cm}

Yuji Kajiyama$^{a,b,c}$\footnote{kajiyama@muse.sc.niigata-u.ac.jp}
,
Hiroshi Okada$^{d,e}$\footnote{HOkada@Bue.edu.eg}
 and
Takashi Toma$^{f,g}$\footnote{t-toma@hep.s.kanazawa-u.ac.jp}
\vspace{5mm}

{\it%
$^{a}${National Institute of Chemical Physics and Biophysics,\\[-1.5mm]
Ravala 10, Tallinn 10143, Estonia}\\
$^{b}${Department of Physics, Niigata University,~Niigata,  950-2181, Japan}\\
$^{c}${Akita Highschool, Tegata-Nakadai 1, Akita, 010-0851, Japan}\\
$^{d}${Centre for Theoretical Physics, The British University in 
Egypt,\\[-1.5mm] El Sherouk City, Postal No, 11837, P.O. Box 43, Egypt}\\
$^{e}${School of Physics, KIAS, Seoul 130-722, Korea}\\
$^{f}${ Institute for Theoretical Physics, Kanazawa University, Kanazawa, 920-1192, Japan}\\
  $^{g}${Max-Planck-Institut f\"ur Kernphysik, Postfach 103980, 69029 Heidelberg, Germany}}

\vspace{8mm}

\abstract{
We study a fermionic dark matter in a non-supersymmetric extension of
 the standard model with a family symmetry based on $D_6 \times
 \hat{Z_2} \times Z_2$. 
In our model, the final state of the dark matter annihilation is
determined to be $e^+ e^-$
by the flavor symmetry, which is consistent with the PAMELA result. 
At first, we show that our dark matter mass should be within the range of $230\ {\rm GeV}- 750\ {\rm GeV}$ in the WMAP analysis combined with $\mu\rightarrow e\gamma$ constraint. Moreover we simultaneously explain the experiments of direct and indirect detection,
by simply adding a gauge and $D_6$ singlet real scalar field. In the direct detection experiments, we show that the lighter dark matter mass $\simeq 230\ {\rm GeV}$ and the lighter standard model 
Higgs boson $\simeq 115\ {\rm GeV}$ is in favor of the observed bounds reported by CDMS II and XENON100. In the indirect detection experiments, we explain the positron excess reported by PAMELA through the Breit-Wigner enhancement mechanism. 
We also show that our model is consistent with no antiproton excess suggested by PAMELA. 
}

\end{center}
\end{titlepage}
\setcounter{footnote}{0}

\section{Introduction}

The existence of the dark matter (DM) in the Universe has been established by measurements. 
The WMAP experiment tells us that the amount of the DM is considered about 23$\%$ of 
energy density of the Universe \cite{wmap}. 
As indirect detection experiments of the DM, PAMELA reported excess of positron fraction in 
the cosmic ray \cite{Adriani:2008zr}. This observation 
can be explained by annihilation and/or decay of DM particles with mass of ${\cal O}(10^{2-3})$ GeV. 
The PAMELA experiment searches antiproton as well in the cosmic ray, and 
it is consistent with the background \cite{Adriani:2008zq}. 
Therefore, if these signals are from annihilation and/or decay processes of DM particles, 
this indicates that the leptophilic DM is preferable. However, even if the DM is leptophilic, 
the resultant positron fraction depends on the generation of final state leptons. 
For instance, if the final state of annihilation and/or decay of the DM is $\tau^+ \tau^-$, 
it will overproduce gamma-rays as final state radiation \cite{strumia}. Therefore it is 
considerable that the leptophilic DM can reflect flavor structure of elementary particles. 
In this point of view, several works discussing the DM and flavor structure have been done 
so far \cite{Adulpravitchai:2011ei,a4-cosmic,Kajiyama:2010sb,Daikoku:2010ew,Hirsch:2010ru,Esteves:2010sh,Meloni:2010sk,Boucenna:2011tj,Kajiyama:2006ww}.    

Flavor structure of elementary particles is thought to be determined by symmetry, 
so called flavor symmetry \cite{review}. 
In our previous work \cite{Kajiyama:2006ww}, we have discussed fermionic DM model with the 
standard model (SM) extension with the $D_6$ flavor symmetry \cite{Dihedral}.  
In this model, three generations of matter fields including right-handed neutrinos are embedded into 
doublet and singlet representations of $D_6$ group in particular way.  
The light neutrino masses are induced by radiative correction through inert $SU(2)_L$ doublet Higgs bosons $\eta$ which do not have vacuum expectation values(VEVs) 
\cite{radseesaw,radiative-seesaw}. We identify a heavy Majorana neutrino 
of $D_6$ singlet $n_S$ with the DM candidate. The DM $n_S$ is stable because of the additional $Z_2$ symmetry. Since the $D_6$ symmetry completely determines flavor structure of the model, 
the final states of annihilation of the DM via Dirac Yukawa interaction $\eta^{\dag} L n_S$ are 
fixed to be electron-positron pair and $\tau$ neutrino pair. 
In that paper, we have found that the DM mass is constrained to be in the range 
$230\ {\rm GeV}- 750\ {\rm GeV}$ from 
the condition of the relic abundance and $\mu \to e \gamma$ process. 
However, this annihilation of the DM via $\eta$-mediated t- and u-channel processes does not 
give enough s-wave contribution to the cross section because it is proportional to 
mass of the final state $m_{e,\nu}$. 
Therefore, this model requires very large enhancement $(\sim 10^6)$ of the cross section at the 
present Universe compared with that at the early Universe 
to explain the PAMELA data, which is not realistic. 

In this paper, we extend our $D_6$ model of Ref.\cite{Kajiyama:2006ww} 
by adding gauge and $D_6$ singlet scalar field $\varphi$, 
which couples with the DM $n_S$ as $\varphi n_S n_S$. 
While the final states of the DM annihilation are the same as those of the previous model, 
there exist s-channel annihilation processes mediated by $\varphi$. In this case, the Breit-Wigner enhancement 
mechanism works which can give enough boost factor \cite{breit,resonance,resonance2}. 
Moreover, the new field $\varphi$ mixes with the $D_6$ singlet Higgs doublet $\phi_S$, 
which is responsible for mass of the quark sector. 
This mixing can simultaneously induce antiproton production by DM annihilation and interaction with quarks 
in atoms. We find that the spin-independent cross section of the DM and quarks 
via the mixing between $\varphi$ and 
$\phi_S$ can be close to sensitivities of direct DM detection experiments such as CDMS II 
\cite{cdms} and XENON100 \cite{xenon10}, 
suppressing antiproton flux in the cosmic ray.


This paper is organized as follows. In section 2, we review our model briefly and summarize the predictions for lepton sector coming from the flavor symmetry. 
In section 3, we analyze the Higgs potential and mixing between the SM Higgs and 
new singlet scalar $\varphi$. In section 4, we show constraints of DM mass from WMAP and 
$\mu \to e\gamma$ process. We discuss direct and indirect detection of DM in section 5 and 6, respectively. 
Section 7 is devoted to conclusions and discussions.

\section{The Model}
In this section, we briefly review a SM extension with $D_6 \times \hat Z_2 \times Z_2$ family symmetry 
\cite{Kajiyama:2006ww}.
\subsection{Yukawa couplings}
We introduce three ``generations" of Higgs doublets $\phi_{I,S}$, inert doublets $\eta_{I,S}$, and one generation of inert singlet $\varphi$. Where $I=1,2$ and $S$ denote $D_6$ doublet and singlet, respectively, and assume that each field is charged in 
specific way under the family symmetry shown in 
Table \ref{a1} and \ref{a2}. 
\begin{table}[thb]
\begin{center}
\begin{tabular}{|c|cccccc|} \hline
 & $L_S$ & $n_S$ & $e^c_S $&$L_I$&$n_I$&$e^c_I$ 
  \\ \hline
 $SU(2)_L\times U(1)_Y$ 
 & $({\bf 2}, -1/2)$  &  $({\bf 1}, 0)$  &  $({\bf 1}, 1)$
 & $({\bf 2}, -1/2)$&  $({\bf 1}, 0)$
 &  $({\bf 1}, 1)$
  \\ \hline
 $D_6$ & ${\bf 1}$  &  ${\bf 1}'''$  &  ${\bf 1}$
 & ${\bf 2}'$&  ${\bf 2}'$&  ${\bf 2}'$
 \\ \hline
 $\hat{Z}_2$
 & $+$ &$+$  & $-$  & $+$ 
 &$+$  & $-$
  \\ \hline
   $Z_2$
 & $+$ &$-$  & $+$  & $+$ &$-$  & $+$ 
   \\ \hline
\end{tabular}
\caption{The $D_6 \times \hat{Z}_2\times Z_2$ 
assignment for the leptons.  The subscript $S$ indicates
a $D_6$ singlet, and the subscript $I$ running from $1$ to $2$
stands for a $D_6$ doublet. $ L_I$ and $L_S$ denote the $SU(2)_L$-doublet leptons,
while $e^c_I$, $e^c_S$, $n_I$ and $n_S$ are the $SU(2)_L$-singlet leptons.
}
  \label{a1}
\end{center}
\end{table}

\begin{table}[thb]
\begin{center}
\begin{tabular}{|c|cccc|c|} \hline
 & $\phi_S$ &$\phi_I$ & $\eta_S$&$\eta_I $ & $\varphi$
  \\ \hline
   $SU(2)_L\times U(1)_Y$ 
 & $({\bf 2}, -1/2)$  &  $({\bf 2}, -1/2)$   &  $({\bf 2}, -1/2)$ 
 & $({\bf 2}, -1/2)$  & $({\bf 1}, 0)$ 
  \\ \hline
 $D_6$ & ${\bf 1}$ &${\bf 2}'$  &  ${\bf 1}'''$  & ${\bf 2}'$ & ${\bf 1}$
 \\ \hline
 $\hat{Z}_2$ &$+$ 
 & $-$ &$+$  & $+$& $+$ 
  \\ \hline
   $Z_2$
 & $+$ &$+$  & $-$  & $-$  & $+$
   \\ \hline
\end{tabular}
\caption{The $D_6 \times \hat{Z}_2\times Z_2$ 
assignment for the Higgs bosons.
}
\label{a2}
\end{center}
\end{table}
Under the $Z_2$ symmetry (which plays the role of $R$ parity
in the MSSM), only the right-handed neutrinos $n_I, n_S$ and 
the inert Higgs doublets $\eta_I, \eta_S$ are odd.
All quarks are assumed to be singlet under the family symmetry so that the 
quark sector is basically the same as the SM, where 
the $D_6$ singlet Higgs doublet $\phi_S$ with
$(+,+)$ of  $\hat{Z}_2\times Z_2$ plays a role in the SM Higgs
in the quark sector. No other Higgs bosons can couple to the quark sector at the tree-level.
In this way we can avoid tree-level flavor changing neutral currents (FCNCs) in the quark sector.
The $\hat{Z}_2$ symmetry is introduced to forbid  tree-level couplings of 
the $D_6$ singlet Higgs $\phi_S$ with $L_I$, $L_S$, $n_I$ and $n_S$, simultaneously to forbid tree-level couplings of $\phi_I, \eta_I$ and $\eta_S$ with quarks. As shall be discussed later, the gauge singlet $\varphi$ plays an important role in explaining an indirect detection reported by PAMELA. 
Furthermore, it is expected to explain the direct detection as CDMS II, because our dark matter $n_S$, 
$D_6$ singlet right-handed neutrino, couples to the quark sector by small mixing between $\varphi$ and 
$\phi_S$, which should be estimated to satisfy the experimental results. 
We will show the numerical analysis of the mixing for both experiments later.


The most general renormalizable $D_6 \times \hat{Z}_2 \times 
Z_2$ invariant 
Yukawa interactions in the lepton sector 
are found to be
\be
{\cal L}_{Y} &=&\sum_{a,b,d=1,2,S}~\left[
Y_{ab}^{ed} (L_{a} i\sigma_2\phi_d) e^c_{b} 
+Y_{ab}^{\nu d} (\eta_d^\dag L_{a}) n_{b} \right]\\
&-& \sum_{I=1,2}\frac{M_1}{2}n_{I}n_{I}-
 \frac{M_S}{2}n_{S}n_{S}
- \sum_{I=1,2}\frac{\mathfrak{S}_1}{2}\varphi n_{I}n_{I}-
 \frac{\mathfrak{S}_S}{2}\varphi n_{S}n_{S}+h.c.,
 \label{wL}
\ee
where the coupling constants $\mathfrak{S}_{1,S}$ are complex in general. 
The electroweak symmetry is broken by the VEVs $\langle \phi_1 \rangle=\langle \phi_2 \rangle\equiv
v_D/2~,\langle \phi_S\rangle=v_S/\sqrt{2},~V^2\equiv v_D^2+v_S^2=(246~{\rm GeV})^2,~\langle \eta_{I,S}\rangle=\langle \varphi\rangle=0$ \cite{okada1}, and we obtain the following mass matrix ${\bf M}_{e} $ and diagonalization matrix $U_{eL}$ of ${\bf M}_{e} {\bf M}_{e}^{\dag} $ in the charged lepton sector:
\be
{\bf M}_{e} = \left( \begin{array}{ccc}
-m^e_{2} & m^e_{2} & m^e_{5} 
\\  m^e_{2} & m^e_{2} &m^e_{5}
  \\ m^e_{4} & m^e_{4}&  0
\end{array}\right),~
U_{eL} &\simeq &\left(
\begin{array}{ccc}
\epsilon_e( 1 - \epsilon_\mu^2) &
-(1/\sqrt{2}) (1-\epsilon_e^2+2\epsilon_e^2 \epsilon_\mu^2) &
1/\sqrt{2}\\
-\epsilon_e( 1 + \epsilon_\mu^2) &
(1/\sqrt{2})(1-\epsilon_e^2-2\epsilon_e^2 \epsilon_\mu^2 ) &
1/\sqrt{2} \\
1-\epsilon_e^2
& \sqrt{2}\epsilon_e &  \sqrt{2} \epsilon_e \epsilon_\mu^2
\end{array}\right),
\label{UeL}
\label{mlepton}
\ee 
where $m^e_{2,4,5}$ are real parameters whose values are determined by observed charged lepton masses $m_{e,\mu,\tau}$. Small parameters $\epsilon_{e,\mu}$ are defined as 
$\epsilon_e=m_e/(\sqrt{2}m_{\mu})$ and $\epsilon_{\mu}=m_{\mu}/m_{\tau}$. In the neutrino sector, Yukawa couplings in the mass eigenstates are given by
\be
Y^S&=&U_{eL}^TY^{\nu S},~
Y^\pm = \frac{1}{\sqrt{2}}U_{eL}^T(Y^{\nu 1}\pm Y^{\nu 2}),\\
Y^S& \simeq &
\left(\begin{array}{ccc}
\!\! 0 \!  &\!  0\!  & \!\!h_{3}\!\! \\
\!\! 0 \!  &\!  0\!  &  \sqrt{2}\epsilon_e h_{3}\!\!  \\
\!\! 0 \!  &\!  0\!  &\!\! 0 \!\! 
\end{array}\right),~\label{yukL}
Y^+ \simeq 
\left(\begin{array}{ccc}
\!\! \frac{h_{4}-2\epsilon_e h_2}{\sqrt{2}}  & 
\!\! \frac{h_{4}}{\sqrt{2}} &  0  \\
\!\! h_2  + \epsilon_e h_{4}  & 
\!\!  \epsilon_e  h_{4}  &  0  \\
\!\! 0  & \!\! h_2  &  0 
\end{array}\right)
\label{yukH},~
Y^- \simeq
\left(\begin{array}{ccc}
\frac{h_{4}}{\sqrt{2}} &
\frac{-h_{4}-2\epsilon_e h_2}{\sqrt{2}}
 &0 \\
\!\!\epsilon_e h_{4} \!\!& 
\!\!h_2  - \epsilon_e h_{4}\!\! &0 \\
\!\!-h_2 \!\!& \!\!0\!\! & 0
\end{array}\right)
\label{yukm}, 
\ee
where the Dirac Yukawa couplings $h_i~(i=2,3,4)$ are of order one. 
Notice that the $D_6$ singlet right-handed neutrino $n_S$ couples only with $L_S$ and $\eta_S$.
Since we consider the case that $\eta_{I,S}$ are inert doublets which do not have VEVs, 
Dirac neutrino mass matrix is not generated and canonical seesaw mechanism does not work. 
Light Majorana neutrino masses are generated by 
radiative seesaw mechanism at one-loop level \cite{radseesaw}. In this mechanism, Majorana mass is 
proportional to $h_i^2 \kappa V^2 M/(16 \pi^2 (M^2-m_{\eta}^2))$, where $\kappa$ denotes typical coupling constant of 
non self-adjoint terms in the Higgs potential. When $\kappa=0$, an exact lepton number $U(1)_{L'}$ invariance is recovered, where the right-handed neutrinos $n_{I,S}$ are neutral under $U(1)_{L'}$ 
in contrast to the conventional seesaw models. This $U(1)_{L'}$ forbids the neutrino masses, so that 
the smallness of the neutrino masses has a natural meaning. 
Now we can derive some predictions of our model based on the family symmetry: 
\begin{enumerate}
\item If $\epsilon_{e,\mu}=0$, the mixing matrix $U_{eL}$ has the maximal mixing in its right-upper block which is the origin of the maximal mixing of atmospheric neutrino mixing. Only an inverted mass spectrum $m_{\nu 3}<m_{\nu 1,2}$ is allowed.
\item  Non-zero $\theta_{13}$ is predicted as $\sin^2 \theta_{13}\simeq \epsilon_e^2=1.2 \times 10^{-5}$. This small value of $\theta_{13}$ is 
consistent with the best fit value $0.020^{+0.008}_{ -0.009}$ with $1 \sigma$ error \cite{valle}.
\item The effective Majorana mass  $\langle m \rangle_{ee}$ is bounded from below as
 $\langle m \rangle_{ee}\gsim0.02~{\rm eV}$.
\end{enumerate}
As a result of this discussion, 
we can assume that $M_{1,S}={\cal O}({\rm TeV})$, $\kappa \ll 1$ and $h_i={\cal O}(1)$. 
Moreover, as can been seen from Eq.(\ref{yukL}), if one identifies the $D_6$ singlet right-handed 
neutrino $n_S$ to be the DM, it mainly couples with electron (and positron) with 
large coupling $h_3 \sim 1$. This selection rule is remarkably determined by the family symmetry.
These facts play a crucial role in the study of cold DM (CDM) as discussed below.

\section{Higgs Potential } 
In this section, we analyze the Higgs potential. 
As discussed in Refs.\cite{Kajiyama:2006ww,okada1}, the Higgs potential 
consists of $D_6$ symmetric and breaking terms. 
Since $D_6$ invariant Higgs potential has an accidental global $O(2)$ symmetry, 
the latter must be introduced in order to forbid massless Nambu-Goldstone (NG) bosons. 
Essentially, such soft $D_6$ breaking terms are mass terms of the Higgs bosons.  
For the potential of $(\phi_I,\phi_S)$, the soft $D_6$  
breaking mass terms \cite{okada1} are given by
\bea
V(\phi)_{soft}=
\mu^2_2(\phi^{\dagger}_2\phi_1+\phi^{\dagger}_1\phi_2)
+\left(\mu^2_4\phi^{\dagger}_S(\phi_1+\phi_2)+h.c.\right), 
\label{softD6break}
\eea
where $\mu^2_2$ is real while  $\mu^2_4$ is complex in general. 
The mass term of $(\phi_I,\phi_S)$ is dominated by Eq.(\ref{softD6break}), and 
subdominantly given by $D_6$ invariant terms of order $V^2$. 
One finds that the $D_6$ breaking terms Eq.(\ref{softD6break}) preserve 
the minimum symmetry $S_2$ under $\phi_1\leftrightarrow\phi_2$.
The key point is that the $S_2$ invariance is required not only to ensure 
the vacuum alignment $\langle \phi_1 \rangle=\langle \phi_2 \rangle \neq 0$ 
but also to forbid NG bosons which violate the electroweak precision test of the SM.

Since the Higgs potential of $\phi_{I,S}$ and $\eta_{I,S}$ are analyzed in 
Ref.\cite{Kajiyama:2006ww}, we do not explicitly show that here again. 
In the present model, the new field $\varphi$ is introduced and 
it plays an important role in our analysis. Therefore we explicitly show 
the potential including $\varphi$. 
The most general renormalizable $D_6 \times \hat{Z}_2 \times 
Z_2$ invariant Higgs potential of $\varphi$ is given by
\be
V(\varphi)&=&m_1^3 \varphi+m^2_2\varphi^2+m_3\varphi^3+\lambda_1\varphi^4,\\
V(\phi,
\varphi)&=&m_4(\phi^{\dagger}_S\phi_S)\varphi+m_5(\phi^{\dagger}_I\phi_I)
\varphi+
\lambda_2(\phi^{\dagger}_S\phi_S)\varphi^2+\lambda_3(\phi^{\dagger}_I\phi_I)\varphi^2,\\
V(\eta, \varphi)&=&V(\phi, \varphi)(\phi\rightarrow \eta),
\ee
where all parameters are considered to be real without loss of generality.
By using the decomposition of $SU(2)_L$ doublets $\phi_{I,S}$, 
\be
\phi_I=\frac{1}{\sqrt2}
\left(\begin{array}{c}
v_D/\sqrt2+\rho_I+i\sigma_I\\
\sqrt2\phi^{-}_I\\
\end{array}\right), ~
\phi_S=\frac{1}{\sqrt2}
\left(\begin{array}{c}
v_S+\rho_S+i\sigma_S\\
\sqrt2\phi^{-}_S\\
\end{array}\right),
\ee
we find the mass matrix 
of neutral Higgs bosons as
\be
H^{t}M^2_hH&=&
\frac{1}{2}\left(
\begin{array}{ccc}
\rho&\sigma&\varphi
\end{array}
\right)
\left(\begin{array}{ccc}
M^2_{\rho,\rho} & M^2_{\rho,\sigma} & M^{2}_{\rho,\varphi}\\
M^2_{\rho,\sigma}  &  M^2_{\sigma,\sigma} & 0\\
M^{2T}_{\rho,\varphi}  &  0 & M^2_{\varphi,\varphi}\\
\end{array}\right)
\left(\begin{array}{c}
\rho\\
\sigma \\
 \varphi\\
\end{array}\right),
\ee 
where $\rho=(\rho_I,~\rho_S)$, $\sigma=(\sigma_I,~\sigma_S)$.
Each $3\times 3$ element $M^2_{\rho,\sigma}$'s are given by \cite{Kajiyama:2006ww}
\be
 M^2_{\rho,\rho}
&\simeq&
\left(\begin{array}{ccc}
0 & 2\mu^2_2 & \sqrt2{\rm Re}(\mu^2_4) \\
2\mu^2_2 & 0 & \sqrt2{\rm Re}(\mu^2_4) \\
\sqrt2{\rm Re}(\mu^2_4) & \sqrt2{\rm Re}(\mu^2_4) &0 \\
\end{array}\right)
+
\left( \begin{array}{ccc}
a_{\rho,\rho} v_D^2& a_{\rho,\rho} v_D^2 &b_{\rho,\rho}
v_D v_S \\ 
a_{\rho,\rho} v_D^2& a_{\rho,\rho} v_D^2&b_{\rho,\rho} v_D
v_S \\ 
b_{\rho,\rho} v_D v_S  &b_{\rho,\rho} v_D v_S
&c_{\rho,\rho} v_S^2\\
\end{array}\right)
,\\
 M^2_{\sigma,\sigma}
&\simeq&
\left(\begin{array}{ccc}
0 & 2\mu^2_2 & \sqrt2{\rm Re}(\mu^2_4) \\
2\mu^2_2 & 0 & \sqrt2{\rm Re}(\mu^2_4) \\
\sqrt2{\rm Re}(\mu^2_4) & \sqrt2{\rm Re}(\mu^2_4) &0 \\
\end{array}\right)\nonumber
\\&&\!\!\!\!\!
+\left( \begin{array}{ccc}
a_{\sigma,\sigma}v_D^2+a'_{\sigma,\sigma}v_S^2&
b_{\sigma,\sigma} v_D^2& c_{\sigma,\sigma}v_D v_S\\
b_{\sigma,\sigma}
v_D^2&a_{\sigma,\sigma}v_D^2+a'_{\sigma,\sigma}v_S^2&c_{\sigma,\sigma}v_D
v_S\\
c_{\sigma,\sigma}v_D v_S&c_{\sigma,\sigma}v_D
v_S&d_{\sigma, \sigma}v_D^2\\
\end{array}\right)
,\\
M^2_{\rho,\sigma}
&\simeq&
\left(\begin{array}{ccc}
0 & 0 & \sqrt2{\rm Im}(\mu^2_4) \\
0 & 0 & \sqrt2{\rm Im}(\mu^2_4) \\
\sqrt2{\rm Im}(\mu^2_4) & \sqrt2{\rm Im}(\mu^2_4) &0 \\
\end{array}\right)
+\left( \begin{array}{ccc}
a_{\rho,\sigma}v_S^2&0&-b_{\rho,\sigma}v_D v_S\\
0&a_{\rho,\sigma}v_S^2&-b_{\rho,\sigma}v_D v_S\\
b_{\rho,\sigma}v_D v_S&b_{\rho,\sigma}v_D v_S&c v_D^2\\
\end{array}\right), 
\ee
where the coefficients $a_{\rho, \rho}$'s are of ${\cal
O}(1)$.
The $\varphi$-dependent terms are given by
\be
\rho M^2_{\rho,\varphi}\varphi&=&
(\rho_1,\rho_2,\rho_S)
\left(\begin{array}{c}
v_Dm_5/\sqrt{2} \\
v_Dm_5/\sqrt{2} \\
v_Sm_4/\sqrt{2}  \\
\end{array}\right)
\varphi,\\
M^2_{\varphi,\varphi}
&=&
2m^2_2+v^2_S\lambda_2+v^2_D\lambda_3.
\ee 
The stable minimum conditions are found by partially differentiating the
potential by 
$\varphi$ as
\be
\left.\frac{\partial V}{\partial \varphi}\right|_{\varphi\rightarrow 0}&=&
m_1^3+\frac{1}{2}\left(v^2_Sm_4+v^2_Dm_5\right)=0,
\ee
and 
\be
\left.\frac{\partial^2 V}{\partial \varphi^2}\right|_{\varphi\rightarrow 0}&=&
M^2_{\varphi,\varphi},~
\left.\frac{\partial^2 V}{\partial \varphi\partial v_{S(D)}}\right|_{\varphi\rightarrow 0}=
\frac{1}{\sqrt{2}}v_{S(D)}m_{4(5)}.
\ee
Therefore, we obtain the vacuum conditions for $\langle \phi_{I,S}\rangle\neq 0$ and 
$\langle \varphi\rangle=0$ as
\be
m_1^3+\frac{1}{2}\left( v_S^2m_4 +v_D^2 m_5\right)=0,~
M^2_{\varphi,\varphi}>0,~v_{S(D)}m_{4(5)}>0. 
\ee
The mass matrix $M^2_h$ is diagonalized by the $7\times 7$ orthogonal matrix ${\cal O}$, as
${\cal O}M^2_h{\cal O}^T$. 
Notice that quarks couple only with $\phi_S$ via Yukawa interactions, and 
$\phi_S(\rho_S)$ mixes with $\varphi$ via $m_4$. 
This mixing parameter $m_4$ will induce both 
interaction of the DM with atoms (direct detection) 
and antiproton flux in the cosmic ray. We will discuss 
these DM phenomenology below. Note also that there is no mixing between 
$\phi$ and $\eta$ because $\eta_{I,S}$ do not get VEVs.

The SM Higgs is described in terms of the linear combination of flavor eigenstate fields as
\bea
SM-Higgs={\cal O}_{11}\rho_1+{\cal O}_{12}\rho_2+{\cal O}_{13}\rho_S+{\cal O}_{14}\sigma_1
+{\cal O}_{15}\sigma_2+{\cal O}_{16}\sigma_S
+{\cal O}_{17}\varphi, 
\eea
and the other combinations correspond to heavy neutral Higgs bosons with 
mass of several hundred GeV. 
Therefore the $\rho_S-\varphi$ mixing is proportional to ${\cal O}^T_
{31}{\cal O}^T_{71}$. In the following analysis, we give numerical values 
of the matrix ${\cal O}$.

\section{WMAP and $\mu\rightarrow e\gamma$ Constraint} 
In this section, we derive conditions for mass of the DM $M_S$ and charged component of 
$\eta$ boson $M_{\eta}$, following the result of Ref.\cite{Kajiyama:2006ww}.  
\subsection{$\mu\rightarrow e\gamma$ Constraint} 
\begin{figure}[t]
\begin{center}
\includegraphics*[width=0.5\textwidth]{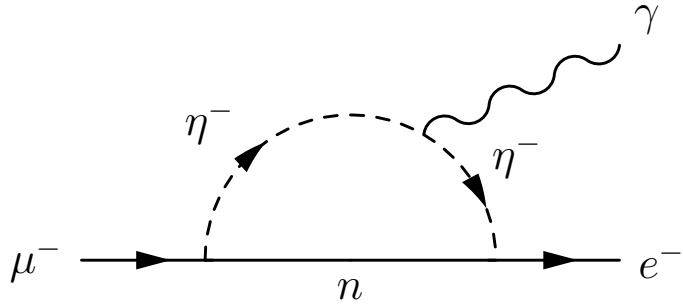}
\caption{The contribution to the $\mu\rightarrow e\gamma$ process.}   
\label{muegamma}
\end{center}
\end{figure}

The DM mass $M_S$ is constrained from the $\mu\rightarrow e\gamma$ process.
The branching fraction 
of $\mu \to e \gamma$ from Fig.\ref{muegamma} is  approximately given by 
\be
B(\mu\to e\gamma)
=\frac{3\alpha}{64\pi G_F^2 } X^4
\simeq \left| X^2 900~\mbox{GeV}^2\right|^2,
 \label{mutoe}
~
X^2 \simeq h_3^2 ~\frac{m_e}{m_\mu}~ \frac{F_2(M_S^2/M_\eta^2)}{M_\eta^2},
\label{mueg}
\ee
and
\be
F_2(x) &=& {1-6x+3x^2+2x^3-6x^2\ln x \over 6(1-x)^4},
\label{func-f2}
\ee
where $x=M_S^2/M_\eta^2$.
As we can see from the Yukawa matrix of Eq.(\ref{yukH}),
only $\eta_S$ couples to  $n_S$ 
with $e_L$ and with $\mu_L$,
where the coupling with $\mu_L$ is suppressed
by  $m_e/m_\mu \simeq 0.005$.
In the next subsection, we will obtain the constraints of the DM mass $M_S$ which 
is consistent with the observed DM relic density $\Omega_d h^2\simeq 0.12$ \cite{wmap}
and $\mu \to e\gamma$, assuming $n_S$ to be the DM. 

\subsection{WMAP} 
In the analysis of Ref.\cite{Kajiyama:2006ww},
we have found that it is more natural and promising that only $n_S$ of three right-handed neutrinos
remains as a fermionic CDM candidate. 
Furthermore since charged component of $\eta_S$ boson couples to $e_L$ and $n_S$ due to our original matrix in Eq.(\ref{yukL}), it remarkably leads to be a clean signal if the charged extra Higgs boson $\eta_S$ is produced at LHC.

\begin{figure}[htb]
\begin{center}
\includegraphics*[width=0.7\textwidth]{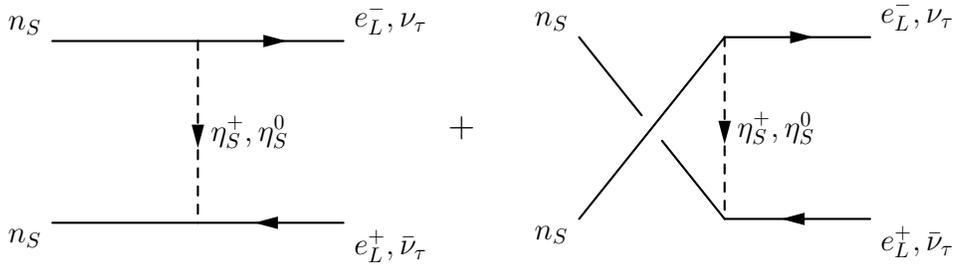}
\end{center}
\caption{\footnotesize
Annihilation diagrams of $n_S$ for the cross section $\sigma_1 v$.
}\label{diagrams}
\end{figure}

We simply find the thermally averaged  cross section $\langle \sigma_1 v \rangle$
for the annihilation of two $n_S$'s \cite{griest1} from Fig.\ref{diagrams} in the limit of the vanishing final state lepton masses:

\be
\langle \sigma_1 v\rangle &=& a_1 + b_1 \frac{6}{x}+\cdots,~
a_1 =0, ~b_1 = \frac{|h_3|^4 r^2 (1-2 r+2 r^2)}{24 \pi M_\eta^2},
\label{bandr}\\
  r &= & M_S^2/(M_\eta^2+M_S^2),\quad x=\frac{M_S}{T}
  \label{randy}
\ee
where $M_\eta$ is $\eta_S$ mass, $M_S$ is $n_S$ mass which is our DM
candidate and $T$ is temperature of the Universe.
The thermally averaged cross section Eq.(\ref{bandr}) does not contain s-wave contribution as a consequence of massless limit of the final state particles, and we find that the allowed region for the DM mass is around ${\cal O}(10^2)\ {\rm GeV}$ from the constraints of WMAP results \cite{wmap} and 
$\mu \to e \gamma$ decay. 

\begin{figure}[htb]
 \begin{center}
\includegraphics*[width=0.5\textwidth]{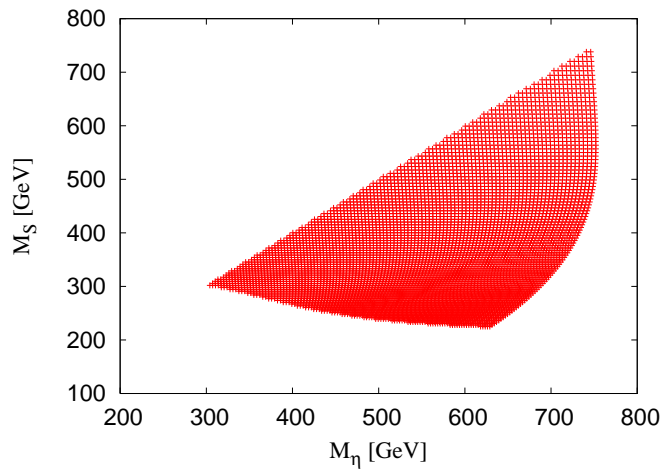}
\end{center}
\caption{\footnotesize
The allowed region in the $M_\eta-M_S$ plane
in which $\Omega_d h^2=0.12,
B(\mu\to e \gamma) < 1.2\times 10^{-11}$ and
$| h_3| <1.5$ are satisfied.
}\label{ms-ms}
\end{figure}

 In Fig.\ref{ms-ms} we present
the allowed region in the $M_\eta-M_S$ plane, in which
$\Omega_d h^2=0.12$
and $B(\mu\to e \gamma) < 1.2\times 10^{-11}$ \cite{pdg} 
are satisfied, where we take $|h_3| <1.5$. 
From Eq.($\ref{bandr}$), retaining $h_3= {\cal O}(1)$ is quite important 
to find the promising DM mass regions, as we mentioned before.
Note that there is no allowed region even for $|h_3|\lsim 0.8$.
As can been seen from Fig.\ref{ms-ms}, we find the mass range as follows:
\be
230\ {\rm GeV}<M_S<750\ {\rm GeV},\quad 300\ {\rm GeV}<M_\eta<750\ {\rm GeV}.
\ee
In this analysis, we have calculated the mass bound for Sunyaev and Zeldovich (SZ) effect \cite{Sunyaev:1980nv}. In our model, $\eta^+_S$, which decays to high energy $e_L^+$, may affect the CMB by the inverse Compton scattering, if the lifetime is not between $10^{-(5-7)}$ sec. From the condition that the lifetime of 
$\eta^+_S$ comes into the allowed region, mass $M_\eta$ has the bound of $30\ {\rm GeV}<M_\eta<750\ {\rm GeV}$.
Where the Yukawa coupling nearly equals to 1, and $M_\eta \gg M_S$ are assumed. Hence, one finds that the SZ effect satisfies the both constraints of $\mu\to e\gamma$ and cosmological pair annihilation of CDMs sufficiently.    

\section{Direct Detection}\label{secdirect}
\begin{figure}[t]
\begin{center}
\includegraphics[scale=1]{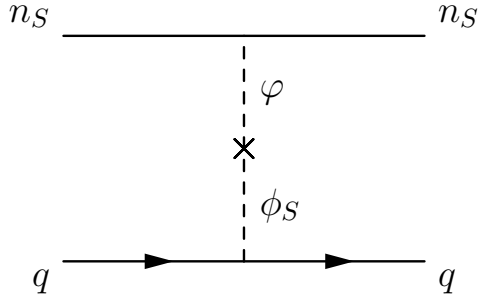}
\caption{The t-channel diagram for the direct detection \cite{cdms}
by the $\varphi-\phi_S$ mixing.}   
\label{dd-diag}
\end{center}
\end{figure}
\begin{figure}[t]
\begin{center}
\includegraphics[scale=0.70]{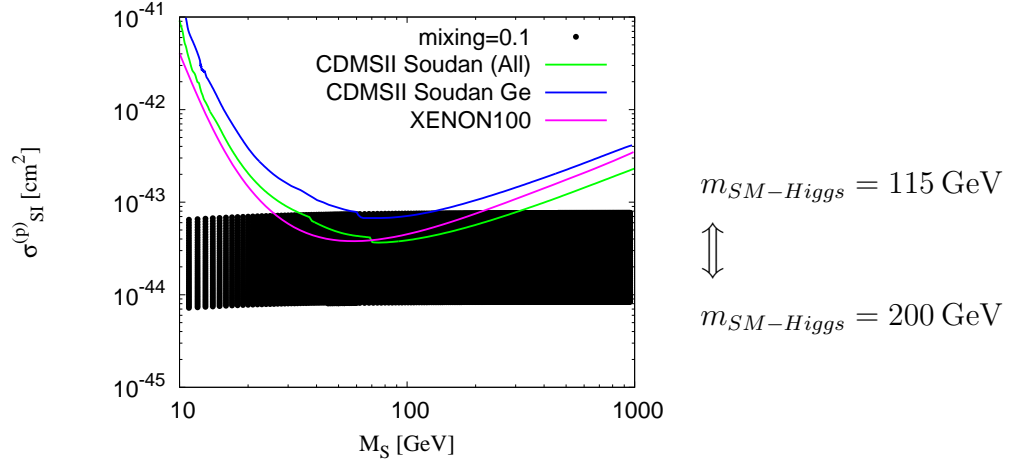}
\put(10,100){${\tiny m_{SM-Higgs}=115\ {\rm GeV}}$}
\put(10,75){${\Big\Updownarrow}$}
\put(10,50){${\tiny m_{SM-Higgs}=200\ {\rm GeV}}$}
\end{center}
\caption{The spin-independent cross section as a function of the DM mass for the direct detection \cite{cdms,xenon10}. The ``mixing" is defined by $|{\cal O}^T_{31}{\cal O}^T_{71}\mathfrak{S}_S|$ and set 
to be $0.1$. The longitudinal black line represents the SM Higgs boson mass range.
}   
\label{cdms}
\end{figure}

We analyze the direct detection search through the experiments of CDMS II \cite{cdms} and XENON100 \cite{xenon10}.
The main contribution to the spin-independent cross section is from the t-channel diagram 
with the mixing between $\varphi$ and $\phi_S$, as depicted in Fig.\ref{dd-diag}. 
Then the resultant cross section for a proton is given by 
\be
\sigma^{(p)}_{SI}=\frac{4}{\pi}\left(\frac{m_p M_S}{m_p+M_S}\right)^2|f_p|^2,
\ee
with the hadronic matrix element
\be
\frac{f_p}{m_p}=\sum_{q=u,d,s}f^{(p)}_{T_q}\frac{\alpha_q}{m_q}
+\frac{2}{27}\sum_{q=c,b,t}f^{(p)}_{TG}\frac{\alpha_q}{m_q},
\ee
where $m_p$ is the proton mass. The effective vertex $\alpha_q$ in our case is given by
\be
\alpha_q\simeq \frac{{\cal O}^T_{31}{\cal O}^T_{71}\mathfrak{S}_SY^q}{m^2_{SM-Higgs}}.
\ee
Here $m_{SM-Higgs}$ is the SM Higgs mass and $Y^q\propto(m_q/v)$ is a Yukawa coupling constant of 
the quark sector. Notice that the quark sector couples only to $\phi_S$.
In the numerical analysis, we set the Higgs masses to avoid the lepton flavor violation (LFV) process 
as follows:
\be
115\ {\rm GeV}\le m_{SM-Higgs}\le 200\ {\rm GeV},~500\ {\rm GeV}\lesssim {\rm other\ six\ 
neutral\ Higgs\ boson\ masses}.
\ee
Under this setup, the elastic cross section is shown in Fig.\ref{cdms}. Where we set ${|\cal O}^T_{31}{\cal O}^T_{71}\mathfrak{S}_S|=0.1$, which we call the ``mixing". 
We plot the DM mass $M_S$ in the region $10-1000$ GeV. 
Since the allowed region of the DM mass is $230\ {\rm GeV}-750\ {\rm GeV}$ from the WMAP analysis combined with $\mu\rightarrow e\gamma$ constraint, rather smaller SM Higgs mass is favored if these experiments could detect the DM near the current bound.

 \section{Indirect Detection}\label{indirect}
The PAMELA experiment implies that there could be positron excess \cite{Adriani:2008zr}, 
but not be antiproton excess \cite{Adriani:2008zq}. In order to describe the PAMELA results successfully through an annihilation process of the DM, we need enhancement of the cross section by using the Breit-Wigner mechanism \cite{breit}. 

\subsection{Positron Production from DM annihilation}
 \begin{figure}[t]
 \begin{center}
\includegraphics*[width=0.4\textwidth]{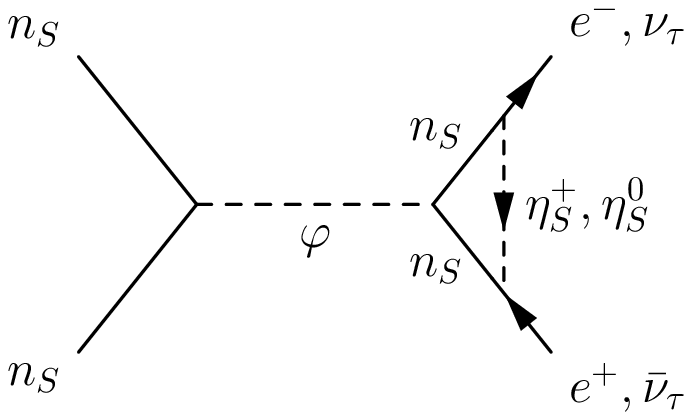}
\end{center}
\caption{\footnotesize
The main process for the positron excess from the DM annihilation. The 
s-channel diagram induces the Breit-Wigner enhancement.
}\label{breit}
\end{figure}
The main channel of the DM annihilation in the present Universe is depicted in Fig.\ref{breit}. 
The $n_S$ annihilation cross section to leptons is given by 
\begin{equation}
(\sigma_2 v)\simeq
\frac{4}{\pi(4\pi)^4}
\frac{m_e^2M_S^4}{M^4_\eta}
\frac{|h_3|^4\left(\mathcal{O}_{R7}\right)^4(\mathrm{Im}\mathfrak{S}_S)^2}
{(s-M^2_R)^2+M^2_R\Gamma^2_R}
\left[(\mathrm{Re}\mathfrak{S}_S)^2\left(I_1^2+I_2^2\right)
+2(\mathrm{Im}\mathfrak{S}_S)^2I_3^2\right],
\label{one-loop}
\end{equation}
\begin{eqnarray}
I_1&=&\int
 d^3x\frac{1-2x_2-2x_3}{(x_1+x_3)\alpha+x_2-x_1x_2\alpha+x_2(x_2-1)\beta}
\delta(x_1+x_2+x_3-1),\\
I_2&=&\int
 d^3x\frac{1-2x_3}{(x_1+x_3)\alpha+x_2-x_1x_3\alpha+x_2(x_2-1)\beta}
\delta(x_1+x_2+x_3-1),\\
I_3&=&\int
 d^3x\frac{1}{(x_1+x_3)\alpha+x_2-x_1x_3\alpha+x_2(x_2-1)\beta}
\delta(x_1+x_2+x_3-1),\\
s&=&E_{\mathrm{cm}}^2\simeq
 4M^2_S\left(1+\frac{v^2}{4}\right),\label{mands}\\
\Gamma_R&=&\frac{\mathcal{O}_{R7}^2}{16\pi}M_R\sqrt{\Delta}
\left[\Delta\left(\mathrm{Re}\mathfrak{S}_S\right)^2+\left(\mathrm{Im}\mathfrak{S}_S\right)^2\right], 
\end{eqnarray}
where the spin of initial states is averaged and $\alpha=M_S^2/M_\eta^2$
and $\beta=m_e^2/M_\eta^2$. Notice that Eq.(\ref{one-loop}) has the s-wave contributions 
because the coupling $\mathfrak{S}_S$ is complex
.
The mass parameter $M_R$ is a mass eigenvalue of the Higgs mass matrix $M_h^2$ 
which is satisfied the resonance mass relation $M_R\simeq
2M_S$\footnote{We take account of physical pole ($\Delta>0$). Unphysical
pole analysis is studied in detail in \cite{breit}}, 
$\Gamma_R$ is the decay width to $n_Sn_S$\footnote{Although there are
other decay channels like $\phi_S\phi_S$, we assume that decay
to $n_Sn_S$ is dominant to lead to the Breit-Wigner enhancement. } and $\Delta=1-4M_S^2/M_R^2$.
The resonance particle $R$ is described in terms of the linear
combination of flavor eigenstate fields as
\begin{equation}
R=\mathcal{O}_{R1}\rho_1+\mathcal{O}_{R2}\rho_2+\mathcal{O}_{R3}\rho_S
+\mathcal{O}_{R4}\sigma_1+\mathcal{O}_{R5}\sigma_2+\mathcal{O}_{R6}\sigma_S
+\mathcal{O}_{R7}\varphi.
\end{equation}
There are the other contributions to the $n_S$ annihilation cross
section such as $t,u$-channel in Fig.\ref{diagrams} or the interference
contributions between $t,u$-channel and $s$-channel.
However all we have to consider is
the contribution of Eq.(\ref{one-loop}) because this is dominant at the
present Universe that the DM relative velocity is $v\sim 10^{-3}$.
One finds that the flavor symmetry remarkably fixes the final states to be positron/electron in our scenario \footnote{We assume that $n_I$ in the loop does not contribute to the positron production because $n_I$ can produce the tauon final state with no suppression, which is now forbidden by the Fermi-LAT $\gamma$-ray experiment \cite{strumia}. Such a condition can be realized in our model by controlling the 
coupling $\mathfrak{S}_1$ to be small. }. 

The thermally averaged annihilation cross section
$\left<\sigma_2 v\right>$ is defined as 
\begin{equation}
\langle\sigma_2v\rangle
\equiv\frac{\int d^3p_1d^3p_2(\sigma_2v)f^{eq}_1f^{eq}_2}{\int d^3p_1d^3p_2f^{eq}_1f^{eq}_2},
\label{num}
\end{equation}
where $p_i$ is the momentum of initial particle $i$ and
 $f_i^{eq}=e^{-E_i/T}$ is the Maxwell-Boltzmann distribution function. 
If we can expand the annihilation cross section in terms of $v^2$ as
$\sigma_2 v=a_2+b_2v^2$, we can calculate it easily as $\left<\sigma_2v\right>=a_2+6b_2/x$, where 
$x=M_S/T$.
Although such a naive treatment is not justified when the annihilation
cross section has a resonance point, an approximate estimation is
 obtained as follows if the condition $\gamma_R\ll \Delta$ is
 satisfied \cite{resonance,resonance2}:
\begin{equation}
\left<\sigma_2v\right>\simeq
\frac{\left(|h_3|^2\left(\mathcal{O}_{R7}\right)^2\mathrm{Im}\mathfrak{S}_S\right)^2}
{(4\pi)^4\pi^{1/2}}
\left[\left(\mathrm{Re}\mathfrak{S}_S\right)^2
\left(\frac{I_1^2}{2}+\frac{I_2^2}{2}\right)
+\left(\mathrm{Im}\mathfrak{S}_S\right)^2I_3^2\right]
\frac{m_e^2}{M_\eta^4}
\frac{\sqrt{\Delta}}{\gamma_R}
x^{3/2}e^{-x\Delta},
\end{equation} 
where $\gamma_R=\Gamma_R/M_R$. Since $\Gamma_R$ is proportional to
$\sqrt{\Delta}$, one might suspect that large annihilation cross section 
is obtained under the condition $\gamma_R/\Delta\ll 1$. We will discuss this point below. 

\begin{figure}[t]
\begin{center}
\includegraphics*[scale=0.95]{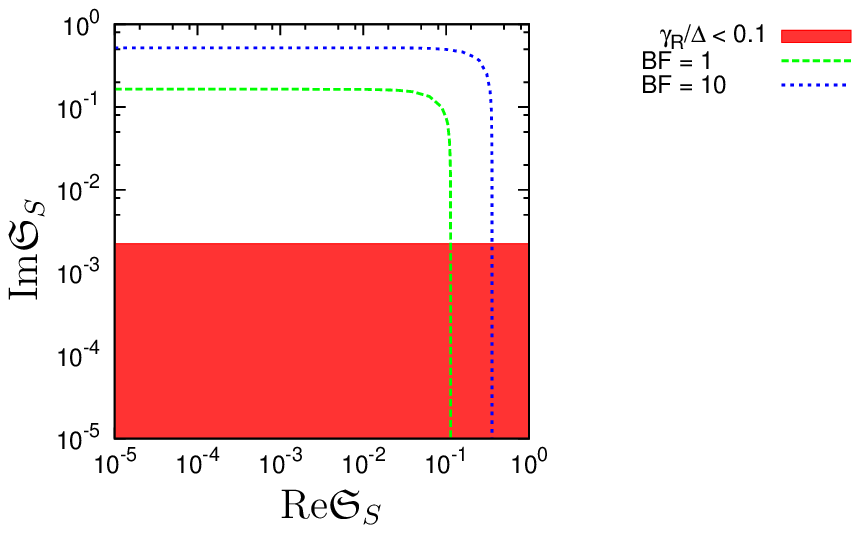}
\qquad\quad
\includegraphics*[scale=0.95]{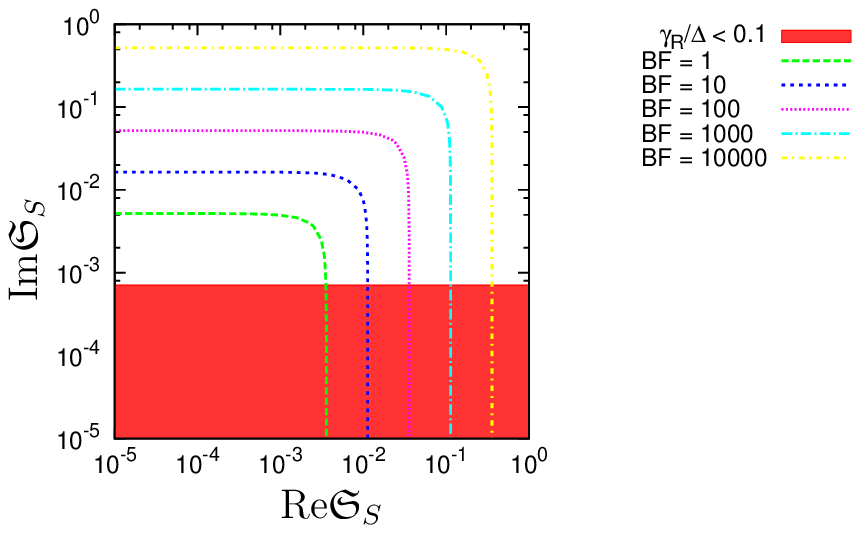}
\end{center}
\caption{\footnotesize{Contours of the boost factor ($BF$) in the
 $\mathfrak{S}_S$ complex plane. 
The parameter $\Delta$ is taken as $10^{-12}$ in the left figure and $10^{-14}$ in the right
 figure.
We set the other parameters as $M_\eta=500$ {\rm GeV},
 $M_S=230\mathrm{GeV}$ and $|h_3|^4\mathcal{O}_{R7}^2=1$.
}}
\label{breit2}
\end{figure}

We define the boost factor $BF$ as
\begin{equation}
BF\equiv \frac{\left<\sigma_2v\right>}{3.0\times
 10^{-9}[\mathrm{GeV}^{-2}]},
\end{equation}
and contours of the boost factor are shown in Fig.\ref{breit2}, where
the red regions satisfy the condition $\gamma_R/\Delta< 0.1$ and
$M_\eta=500$ GeV is taken as a typical example.
The degree of the fine tuning is smaller(larger) if the smaller(larger) $M_\eta$ value
is taken because the thermally averaged cross section
$\left<\sigma_2v\right>$ is inversely proportional to $M_\eta^4$.
One finds that a large boost factor is obtained through the
Breit-Wigner enhancement from Fig.\ref{breit2} if the parameters
satisfy 
$\Delta\lesssim 10^{-13}$ and $\mathrm{Im}\mathfrak{S}_S\ll\mathrm{Re}\mathfrak{S}_S$. 
Under this condition, the thermally averaged annihilation cross section
and the decay width of $R$ are written as
\begin{eqnarray}
\left<\sigma_2 v\right>&\simeq&
\frac{16\sqrt{\pi}}{(4\pi)^4}|h_3|^4\mathcal{O}_{R7}^2\left(\mathrm{Re}\mathfrak{S}_S\right)^2\left(\frac{I_1^2}{2}
+\frac{I_2^2}{2}\right)\frac{m_e^2}{M_\eta^4}x^{3/2}e^{-x\Delta},\\
\Gamma_R&\simeq&
\frac{\left(\mathrm{Im}\mathfrak{S}_S\right)^2}{16\pi}\mathcal{O}_{R7}^2M_R\sqrt{\Delta},
\end{eqnarray}
thus one find that $\mathrm{Im}\mathfrak{S}_S\ll\mathrm{Re}\mathfrak{S}_S$ is important to obtain large annihilation cross section and small
decay width.

The flux of positron and electron from DM annihilation is given by
$\Phi_{e^\pm}(\epsilon)$ \cite{Hisano:2005ec} and the positron fraction is given by
\be
{\rm Positron~Fraction}\equiv \frac{\Phi_{e^+}(\epsilon)+\Phi^{\rm{sec.}}_{e^+}(\epsilon)}
{\Phi_{e^+}(\epsilon)+\Phi^{\rm{sec.}}_{e^+}(\epsilon)
+\Phi_{e^-}(\epsilon)
+\Phi^{\rm{prim.}}_{e^-}(\epsilon)+\Phi^{\rm{sec.}}_{e^-}(\epsilon)},
\ee
where $\Phi_{e^\pm}(\epsilon)$ are the contributions from DM
annihilation, and the others 
are the background fluxes given by 
\be
\Phi^{\rm{prim.}}_{e^-}(\epsilon)
&=&
\frac{0.16 \epsilon^{-1.1}}{1+11 \epsilon^{0.9}+3.2 \epsilon^{2.15}}(\rm{GeV}^{-1}{\rm cm}^{-2}\rm{s}^{-1}\rm{sr}^{-1}),
\nn\\
\Phi^{\rm{sec.}}_{e^-}(\epsilon)
&=&
\frac{0.70 \epsilon^{0.7}}{1+110 \epsilon^{1.5}+600 \epsilon^{2.9}+580 \epsilon^{4.2}}(\rm{GeV}^{-1}{\rm cm}^{-2}\rm{s}^{-1}\rm{sr}^{-1}),
\nn\\
\Phi^{\rm{sec.}}_{e^+}(\epsilon)
&=&
\frac{4.5 \epsilon^{0.7}}{1+650 \epsilon^{2.3}+1500\epsilon^{4.2}}(\rm{GeV}^{-1}{\rm cm}^{-2}\rm{s}^{-1}\rm{sr}^{-1}).
\ee

 \begin{figure}[htb]
 \begin{center}
\includegraphics*[width=0.5\textwidth]{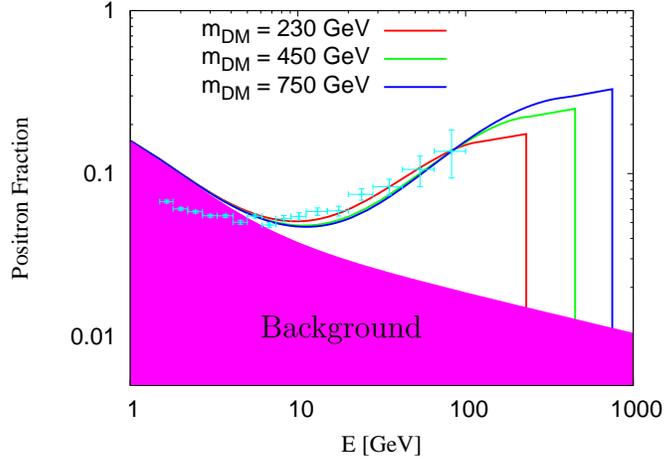}
\put(-160,45){ Background}
\end{center}
\caption{\footnotesize
The positron fraction for DM annihilation($n_S$) into $e^+e^-$. 
The red, green and blue lines are the best fits for $M_S=230~{\rm GeV}$ and $\langle \sigma_2 v \rangle =8.5\times 10^{-8}~
{\rm GeV}^{-2}$, $M_S =450~{\rm GeV}$ and $\langle \sigma_2 v\rangle =
  2.6 \times10^{-7}~{\rm GeV}^{-2}$ and $M_S =750~{\rm GeV}$ and
  $\langle \sigma_2 v\rangle = 6.8\times10^{-7}~{\rm GeV}^{-2}$ respectively.
}\label{fraction}
\end{figure}
The direct positron fraction is plotted in Fig.\ref{fraction} for some
fixed parameters.
The $BF$ of order $10^2$ is required in all cases. This $BF$ is not large enough to fit the Fermi-LAT data
\cite{fermi}.
Thus the constraints from diffuse gamma rays and
neutrinos are not severe as long as isothermal dark matter profile is
considered \cite{strumia}.
It might be worth mentioning that the DM mass less than ${\cal O}\ ({\rm TeV})$ is in favor of the experiment recently reported by HESS \cite{Abramowski:2011hc}, 
if one considers the NFW profile \cite{Navarro:1996gj}.

\subsection{Muon Flux Measurement from Super-Kamiokande}
We briefly mention that the high energy neutrinos induced by DM annihilations in the earth, the sun, 
and the galactic center are an important signal for the indirect detection of the DM \cite{Hisano:2008ah}. 
Such energetic neutrinos induce upward through-going muons from charged
current interactions, which provide the most effective signatures in Super-Kamiokande (SK) \cite{Desai:2004pq}. 
Once the thermally averaged cross section of the muon flux reaches the same order of the cross section required by the PAMELA results, 
it is natural to expect that such a value of cross section is close to 
the upper bound of the muon flux measured by SK.
In fact, our model has the large cross section enhanced by the Breit-Wigner mechanism with $\nu_{\tau}$ pair final state as can been seen in Fig.\ref{breit}. 
However since the total cross section is proportional to the neutrino mass as in Eq.(\ref{one-loop}),
the neutrino flux is extremely suppressed than positron and electron 
fluxes. 

\subsection{Antiproton Production from DM Annihilation}
 \begin{figure}[htb]
 \begin{center}
\includegraphics*[width=0.5\textwidth]{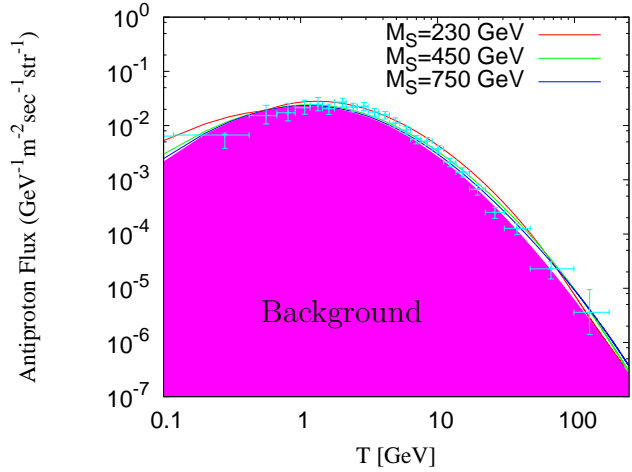}
\put(-150,55){Background}
\end{center}
\caption{\footnotesize
The antiproton flux is mainly induced by the top-pair production through the Higgs mixing between $\varphi$ and $\phi_S$ in our model. Where $T=E-M_{S}$ is the kinetic energy of antiproton, we set $\langle\sigma_3 v\rangle=3\times 10^{-9}$ ${\rm GeV^{-2}}$.
}\label{antiproton}
\end{figure}
Finally, we briefly discuss the antiproton flux in the cosmic ray. Since our model has
the quark-DM coupling through the Higgs mixing between $\varphi$ and
$\phi_S$, as discussed in section \ref{secdirect}, we
have to verify that our antiproton flux is consistent with
the antiproton experiment of PAMELA. The main source comes from
the top quark pair production, and substantially bottom and charm pair
production. 
The cross section of the $n_Sn_S\rightarrow q\bar q$ processes is given by
\be
\sigma_3v(n_Sn_S\rightarrow q\bar q)&=&\frac{1}{2\pi
}\sqrt{1-\frac{m^2_q}{M_S^2}}\frac{m_q^2}{v_S^2}
\frac{(\mathrm{Im}\mathfrak{S}_S)^2M_S^2}{(s-M_R^2)^2+M_R^2\Gamma_R^2}\nonumber\\
&&\times\left[
\left(\mathcal{O}_{R3}\mathcal{O}_{R7}\right)^2\left(1-\frac{m_q^2}{M_S^2}\right)
+\left(\mathcal{O}_{R6}\mathcal{O}_{R7}\right)^2
\right],
\ee
where the index $q$ is summed over top, bottom, and charm
quark. The energy-squared $s$ of the initial state is defined in Eq.(\ref{mands}). 

The thermally averaged annihilation cross section is expressed in terms of
$\Delta$, $\gamma_S$ and some couplings as
\begin{equation}
\left<\sigma_3v\right>\simeq
\frac{\left(\mathrm{Im}\mathfrak{S}_S\right)^2}{16\pi^{1/2}}
\sqrt{1-\frac{m_q^2}{M_S^2}}\frac{m_q^2}{v_S^2M_S^2}
\left[
\left(\mathcal{O}_{R3}\mathcal{O}_{R7}\right)^2\left(1-\frac{m_q^2}{M_S^2}\right)
+\left(\mathcal{O}_{R6}\mathcal{O}_{R7}\right)^2
\right]
x^{3/2}
\frac{\sqrt{\Delta}}{\gamma_R}e^{-\Delta x}.
\end{equation}
The PAMELA experiment implies the positron excess, but no antiproton
excess. Thus the ratio of the annihilation cross section to leptons and
quarks constrains the mixing parameters between $\varphi$ and $\phi_S$.
The ratio is given by 
\begin{equation}
R\equiv\frac{\left<\sigma_3v\right>}{\left<\sigma_2v\right>}
\sim
\left(\frac{m_q}{m_e}\right)^2\left(\frac{M_\eta^4}{v_S^2M_S^2}\right)
\frac{(4\pi)^4}{|h_3|^2}\frac{\mathcal{O}_{R3}^2+\mathcal{O}_{R6}^2}
{\left(\mathrm{Re}\mathfrak{S}_S\right)^2+\left(\mathrm{Im}\mathfrak{S}_S\right)^2},
\end{equation}
where $I_1$, $I_2$ and $I_3$ are taken as $\mathcal{O}(10^{-2})$ which
is evaluated by numerical analysis.
If we require the boost factors for leptons and quarks to be 100 and 1
respectively, the constraint to the couplings becomes 
\begin{equation}
\frac{\mathcal{O}_{R3}^2+\mathcal{O}_{R6}^2}
{\left(\mathrm{Re}\mathfrak{S}_S\right)^2+\left(\mathrm{Im}\mathfrak{S}_S\right)^2}
\lesssim \mathcal{O}(10^{-24}),
\end{equation}
where we have taken the masses of  $M_S=450$ GeV and $M_\eta=500$ GeV.
We find that the mixing matrix elements $\mathcal{O}_{R3}$ and $\mathcal{O}_{R6}$ which appear in
$\left<\sigma_3v\right>$ need to be suppressed by $\mathcal{O}(10^{-12})$ in order to have no antiproton excess
if $\mathfrak{S}_S$ is $\mathcal{O}(1)$.

The flux of antiproton from DM annihilation is given in Ref.\cite{Hisano:2005ec}.
We plot the antiproton flux as a function of the kinetic energy of antiproton $T=E-M_{S}$ in Fig.\ref{antiproton}. Where we adopt $\langle\sigma_3 v\rangle=3\times 10^{-9}$ ${\rm GeV^{-2}}$, {\cal i.e.} $BF=1$, which is required to explain the WMAP experiment, and the same set up as the positron case.
The key parameters contributing to the direct detection are $\mathcal{O}_{13}$ and $\mathcal{O}_{17}$ which come from SM Higgs mediation, while those to the indirect detection of the antiproton are $\mathcal{O}_{R3}$, $\mathcal{O}_{R6}$, and $\mathcal{O}_{R7}$ which come from resonant bosons. It suggests that both of them can be explained by independent way.
Hence it is easy to find the allowed region avoiding such an enhancement as well by controlling many parameters in the Higgs sector.

\section{Summary and Conclusions}
In this paper, we have considered that two important issues of the dark matter in a non-supersymmetric extension of the radiative seesaw model with a family symmetry based on $D_6 \times \hat{ Z}_2 \times Z_2$: direct detection recently reported by CDMS II and indirect detection  reported by PAMELA.
We suppose that the $D_6$ singlet right-handed neutrino is the promising candidate of the DM. Analyzing the $\mu \to e \gamma$ together with the WMAP result, we have shown the allowed region for the DM mass to 
be $230~{\rm GeV}<M_S<750~{\rm GeV}$, within a perturbative regime. 
In the analysis of the direct detection experiment of CDMS II and XENON100, we have shown that the Higgs mixing between $\varphi$ and $\phi_S$ plays an important role in generating the quark effective couplings, and also there exist allowed region to be detected by those experiments in near future.
As a result of the positron production analysis through PAMELA, a couple of remarks are in order.
In the case of $M_S=230~{\rm GeV}$, $M_S=450~{\rm GeV}$ and $M_S=750~{\rm GeV}$, each of $\langle \sigma v\rangle = 8.5 \times10^{-8}~{\rm GeV}^{-2}$, $\langle \sigma v\rangle = 2.6\times10^{-7}~{\rm GeV}^{-2}$ and $\langle\sigma v\rangle = 6.8\times10^{-7}~{\rm GeV}^{-2}$ is required, respectively. 
In all cases the required boost factor is at most $\sim\mathcal{O}(10^2)$, which is realized by the 
Breit-Wigner enhancement mechanism if the sutable parameter regions are choosen. 
Also such boost factor is not large enough to fit the Fermi-LAT result. 
Thus constraints from diffuse gamma rays and
neutrinos are not severe as long as isothermal dark matter profile is considered.
Finally, we have investigated the antiproton flux in the cosmic ray to compare to the direct detection. 
We found that the constraint of the mixing from the direct detection
can easily satisfy the allowed region for no antiproton excess by controlling many parameters in the Higgs sector.

\section*{Acknowledgments}
This work is supported by the ESF grant No.~8090 (Y.K.) and 
Young Researcher Overseas Visits Program for Vitalizing Brain Circulation Japanese in JSPS 
(Y.K. and T.T.). 
H.O.\ acknowledges
partial supports from the Science and Technology Development Fund
(STDF) project ID 437 and the ICTP project ID 30.


\end{document}